\documentclass{article}
\usepackage{amssymb}
\usepackage[english]{babel}
\usepackage[T1]{fontenc}
\usepackage{array}

\usepackage[a4paper,top=3cm,bottom=2cm,left=3cm,right=3cm,marginparwidth=1.75cm]{geometry}
\usepackage{amsmath}
\usepackage{graphicx}
\usepackage[colorinlistoftodos]{todonotes}
\usepackage{cite}
\usepackage[colorlinks=true, allcolors=blue]{hyperref}
\usepackage{authblk}
\usepackage{lmodern}
\usepackage{amsmath,amssymb,amsthm,bm,bbm}
\usepackage{siunitx}
\usepackage{xspace}
\usepackage{graphicx}
\usepackage{hyperref}
\usepackage{braket}
\usepackage{physics}       
\usepackage{enumitem}      
\usepackage{times}
\usepackage{siunitx}  
\usepackage{listings}
\usepackage{xcolor} 
\usepackage{mathtools }
\usepackage{tcolorbox}
\usepackage{orcidlink}

\begin{document}

\title{The Effect of Geometry on Thermodynamic Response}

\author[a,*]{Bojana Bokic}
\author[b,c,d,e*,+]{S\'{e}bastien R. Mouchet\orcidlink{0000-0001-6611-3794}}
\author[a,*]{Biljana Stankov}
\author[f]{Sanja Ostojic}
\author[b,g]{Nicolas Roy}
\author[a]{Darko Vasiljevic}
\author[h]{Yin Chang}
\author[a]{Marija Radmilovic-Radjenovic}
\author[a]{Branislav Radjenovic}
\author[i,+]{Thierry Verbiest}
\author[j,+]{Mohamed Hatifi \orcidlink{0009-0005-3368-2751}}
\author[a,e,+]{Branko Kolaric}

\affil[a]{Photonics Center, Institute of Physics, University of Belgrade, Pregrevica 118, 11080 Belgrade, Serbia}

\affil[b]{Department of Physics \& Namur Institute of Structured Matter (NISM), University of Namur, Rue de Bruxelles 61, 5000 Namur, Belgium}

\affil[c]{Institute of Life, Earth and Environment (ILEE), University of Namur, Rue de Bruxelles 61, 5000 Namur, Belgium}

\affil[d]{School of Physics, University of Exeter, Stocker Road, Exeter EX4 4QL, United Kingdom}

\affil[e]{Micro- and Nanophotonic Materials Group \& Research Institute for Materials Science and Engineering, University of Mons, Belgium}

\affil[f]{Institute of General and Physical Chemistry, Studentski trg 12/V, 11158 Belgrade, Serbia}

\affil[g]{Namur Institute for Complex Systems (naXys), University of Namur, Rue de Bruxelles 61, 5000 Namur, Belgium}

\affil[h]{Department of Materials and Optoelectronic Science, National Sun Yat-sen University,
No. 70, Lienhai Rd., Kaohsiung 80424, Taiwan (R.O.C.)}

\affil[i]{Molecular Imaging and Photonics, Department of Chemistry, KU Leuven, Celestijnenlaan 200D, 3001 Heverlee, Belgium}

\affil[j]{Aix Marseille Univ, CNRS, Centrale M\'editerran\'ee, Institut Fresnel, Marseille, France}

\affil[*]{Co-shared first authorship}
\affil[+]{Co-shared corresponding authorship}
\date{}
\maketitle

\begin{abstract}
 At the nano- and microscale, various patterns influence and shape thermal and optical response, making them essential for the survival of various biological species. In addition, controlling thermal radiation is vital for a broad range of applications, such as thermal management, spectroscopy, optoelectronics, and energy conversion technologies. For this reason, there is strong pressure to elucidate the physics of thermal radiation at the nanoscale. In this article, we provide evidence that complex nanoscale geometries affect thermal management, leading to an unusual thermal response in heat-capacity measurements as a function of temperature. Beyond identifying the structural constraints associated with this unusual thermodynamic response, the current study introduces the possibility of shaping the apparent heat-capacity response through geometry without necessarily altering the system's chemistry.
\end{abstract}

\section{Introduction}

\begin{figure}
\centering
\includegraphics[width=0.9\textwidth]{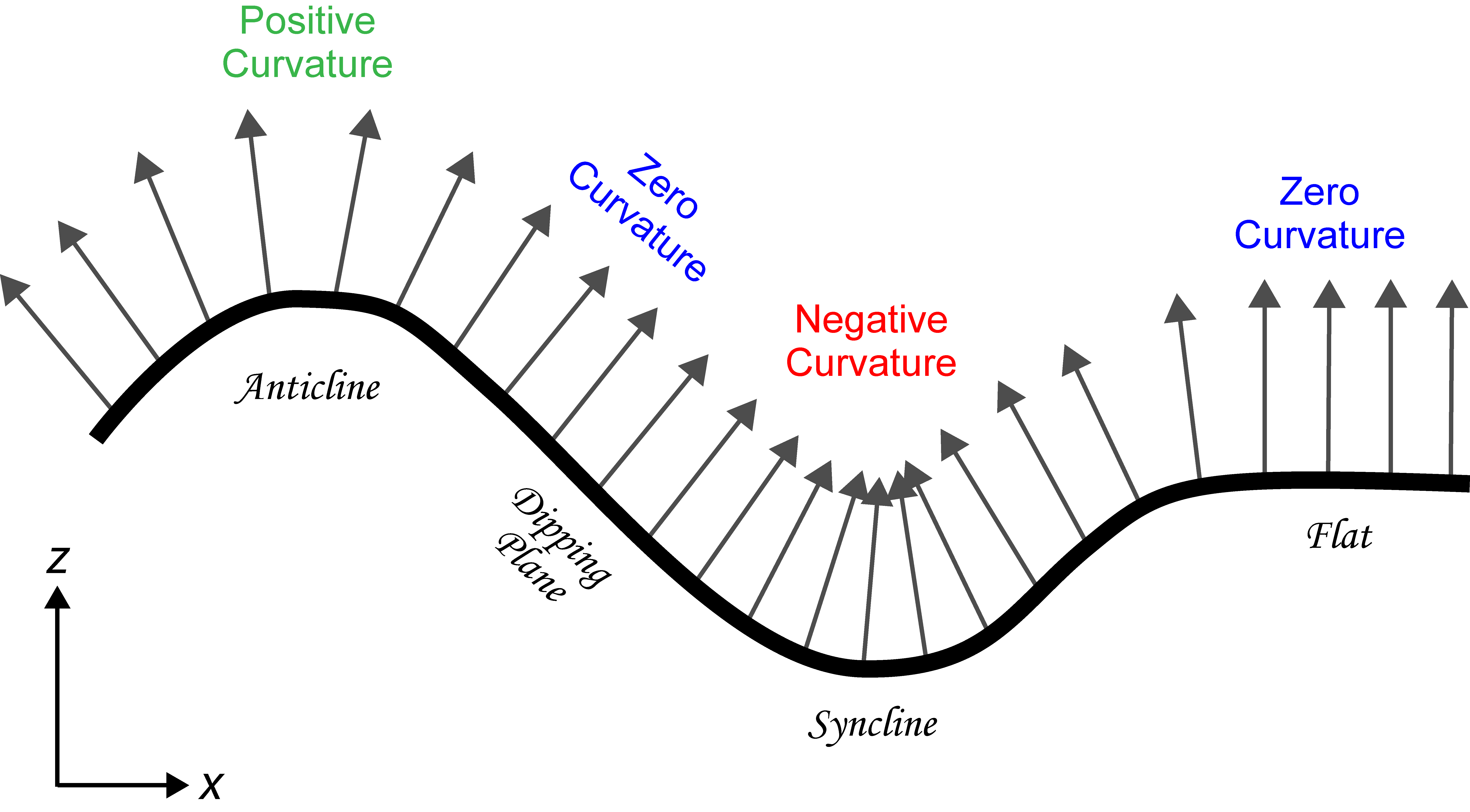}
\caption{\label{fig:ChemPot}Curvature influences the mutual interaction between a surface and its environment. The chemical potential is therefore modified: a positive curve (also referred to as anticline or convex surface) corresponds to a high chemical potential, whereas a negative curvature (also known as syncline or concave surface) corresponds to a low chemical potential.}
\end{figure}

Thermodynamics and optics are two fundamental pillars of classical physics. Although thermodynamics does not describe the "ultimate" nature of reality, it has historically played a key role in shaping atomic and quantum theory~\cite{krumm2017thermodynamics}. At first glance, photonics may appear disconnected from thermodynamics. However, from a broader perspective, the assembly of matter at the nanoscale, including the design of photonic structures, is deeply governed by thermodynamic principles. A notable example is the periodic modulation of the refractive index in photonic crystals achieved through self-assembly methods~\cite{gonzalez2012linear,MADANU2023290,LOURDUMADANU2023233,huang2024colloidal}, enabling both active and passive control of light. Moreover, altering the refractive index influences intermolecular and surface interactions, affecting properties such as wetting~\cite{dellieu2015quantum,israelachvili2011intermolecular,butt2003physics}. More broadly, the ability to control forces between macroscopic bodies or surfaces has become increasingly critical for applications ranging from nanomechanics to the colloidal stability~\cite{butt2003physics,deGennes1985wetting,alhambra2014casimir}.
The thermodynamics of nanostructured materials, including natural photonic materials, differs significantly from conventional bulk thermodynamics~\cite{biro2003,Carrascal2017,Mouchet2025}. Interfaces profoundly influence all thermodynamic parameters, often increasing internal energy compared to the one in bulk~\cite{fultz2016phase}. The role of interfacial entropy in phase stability, mixing, colloidal behavior, and wetting has been explored extensively~\cite{israelachvili2011intermolecular,butt2003physics}. Interestingly, these considerations echo broader questions about whether classical physics alone suffices to explain complex systems such as living matter. Paul Davies recently raised this issue by questioning whether “new physics lurks inside living matter” and suggesting that life might involve principles beyond standard thermodynamics and quantum mechanics~\cite{Davies2020}. While our focus remains on nanostructured photonic materials, this perspective underscores the need to revisit foundational concepts when addressing systems with emergent behavior.
For instance, in a composite nanomaterial consisting of two phases, the change in Helmholtz free energy $F$ for phases ($\alpha$, $\beta$) is expressed as: 
\begin{equation}
d F=-S d T-P d V+\sum \mu_i d N_i+d W.
\end{equation} 
The standard thermodynamic variables describe how energy flows through the system. Here, $S$, $T$, $P$, $V$, $\mu_i$, $N_i$, and $W$
represent the entropy, temperature,
pressure, volume, chemical potential of species $i$,
number of particles of species $i$, and work performed on the system,
respectively.
This expands to: 
\begin{equation}
\begin{aligned}
d F & =-S d T-P^\alpha d V-\left(P^\beta-P^\alpha\right) d V^\beta \\
& +\sum \mu_i^\alpha d N_i^\alpha+\sum \mu_i^\beta d N_i^\beta+\sum \mu_i^\sigma d N_i^\sigma+\gamma d A,
\end{aligned}
\end{equation}
where $P^\alpha$, $P^\beta$, $V^\beta$, $\mu_i^\sigma$, $N_i^\sigma$, $\gamma$, and $A$
denote the pressure in phase $\alpha$, the pressure in phase $\beta$, the volume of phase $\beta$,
the chemical potential of species $i$ at the interface, the number of particles of species $i$ at the interface, the interfacial energy (i.e., surface tension of the interface) and the interfacial area,
respectively. The sum runs over all chemically different components ($i$, $\alpha$, $\beta$, and $\sigma$). The chemical potential $\mu_i$ of the $i^{th}$ component is a key parameter governing energy exchange during physicochemical processes that alter system composition. The chemical potential is one of the most crucial parameters used to describe the properties of a condensed system. Beyond compositional changes, chemical potential also depends on geometry, as curvature modulates distances and surface interactions (Fig.~\ref{fig:ChemPot}). Equation~\ref{eq:ChemicalPotential} explicitly links chemical potential to curvature, revealing the geometric basis for thermodynamic control~\cite{butt2003physics}:
\begin{equation}
\Delta \mu=\gamma \Omega\left(\frac{1}{R_1}+\frac{1}{R_2}\right),
\label{eq:ChemicalPotential}
\end{equation}
where $\Delta\mu$, $\gamma$, $\Omega$, $R_1$, and $R_2$
denote the change in chemical potential, the interface energy, the molecular volume, and the principal radii of curvature, respectively.

For convex surfaces, also called anticlines (Fig.~\ref{fig:ChemPot}), curvature is positive, resulting in a higher chemical potential compared to flat surfaces, whereas concave surfaces (i.e., synclines) exhibit a lower chemical potential. This geometric effect explains why nanomaterials often display properties distinct from bulk materials: surface atoms, with reduced coordination and unsatisfied bonds, are less stabilized and respond differently. More generally, geometry can act as a control parameter for physical dynamics, including in quantum systems where spatial configuration can govern environmental memory and the resulting quantum correlations. Recently, Bravetti proposed a geometrical foundation for thermodynamics~\cite{Bravetti2019Contact}, emphasizing that the energy and entropy of a composite system are additive across its constituent subsystems and independent of the system's size.

In this framework, $S(U,V,N)$ is continuous and differentiable and monotonically increasing in energy. However, the presence of a physical interface (rather than an ideal Gibbs interface), such as in nanostructured materials, can break these ideal assumptions, leading to discontinuities and size-dependent behavior~\cite{Bravetti2019Contact}.
Natural photonics explores how light in the visible and infrared regimes interacts with structures evolved in nature~\cite{mouchet2018structural,Mouchet2021,Mouchet2025}. Such photonic architectures occur in a wide range of organisms, including mammals (e.g., blue eyes in primates), marsupials, fish, birds (e.g., hummingbirds and pigeons), and insects (e.g., butterflies and beetles)~\cite{Verstraete2019,Mara2022}. These natural materials indicate strikingly complex patterns that offer inspiration for next-generation photonic and thermodynamic designs~\cite{Mouchet2025,Delmote2026} and were used as a model in the current study.
Here, we experimentally and theoretically reveal, for the first time, the profound influence of nanostructure geometry and surface corrugation on thermal management. Our results show that the intricate architecture of natural photonic structures with rich and complex interfaces significantly shapes the thermodynamic response of composite systems. Interestingly, in certain cases, these systems surprisingly exhibit a decrease in apparent heat capacity ($c^{\mathrm{app}}_{\mathrm{p}}$) upon heating, which indicates enhanced thermal dissipation mechanisms. The heat capacity quantifies the amount of energy required to increase the temperature of a material by 1°C~\cite{Barron1999HeatCapacity}. This counterintuitive phenomenon highlights the potential of geometry-driven thermodynamic control and opens new avenues for designing advanced materials with customized thermal properties.

\section{Materials and Methods}

\subsection{Samples}

Air-dried specimens of \textit{Morimus asper funereus} and \textit{Euchroma giganteum} beetles were kindly provided by the Biophotonics Laboratory, Institute of Physics, University of Belgrade.  No further sample preparation was necessary for the measurements described, aside from electron microscopy (see next section). All analyses are performed using insect elytra.

\subsection{Morphological characterization}
A field emission gun scanning electron microscope (FEGSEM) (Mira System, TESCAN, Czech Republic) is used for ultrastructural analysis of specimens of \textit{M. asper funereus}. Before analysis, insect elytra are removed and placed on an aluminum mount and coated with a thin layer (5-10 nm) of AuPd (AuPd), using a SC7620 Mini Sputter Coater (Quorum Technologies Ltd., UK). The images were taken at a working distance of about 4~mm.

\subsection{Thermodynamic Measurements}
Differential Scanning Calorimetry (DSC) analysis is performed using a differential scanning calorimeter DSC Q1000, TA Instruments, Delaware, USA. The heat capacities of the samples in the temperature range from 20°C to 95°C or 140°C are determined according to the manufacturer's protocol and in accordance with the standard test method ASTM E1269-11. The maximum temperature is set below 200°C, at which chitin begins to melt.

\subsection{Laser illumination experiment}

\begin{figure}
\centering
\includegraphics[width=1\textwidth]{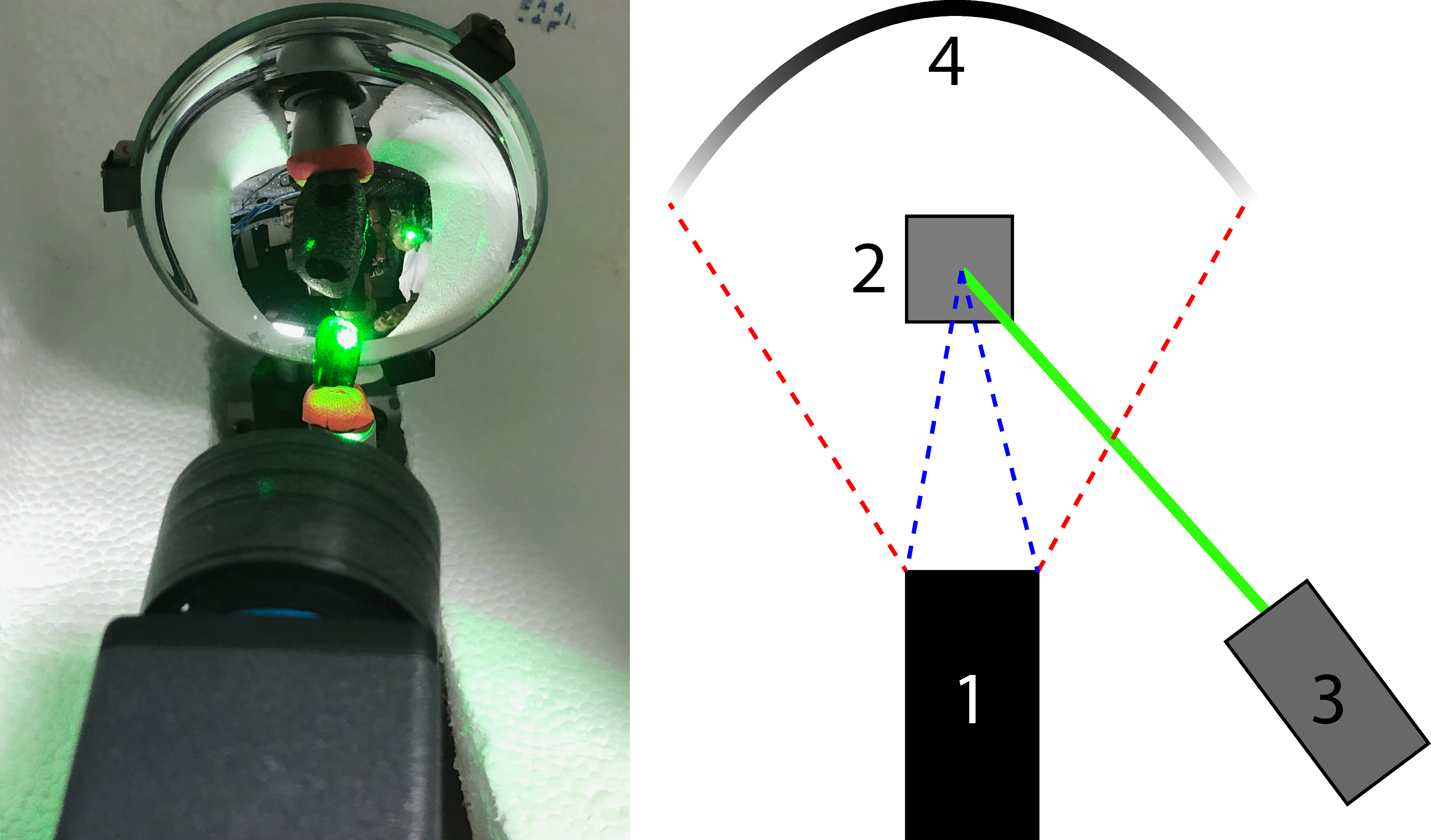}
\caption{The setup is designed to capture both direct and mirror‑reflected thermal images in real time, allowing precise quantification of front and back temperature differences under laser irradiation: (left) setup in the laboratory during the experiment, (right) scheme of the setup (1 - FLIR camera; 2 - aluminum mount with the sample; 3 - laser pointer with a 532-nm wavelength; 4 - spherical mirror).} \label{fig:setup}
\end{figure}

An FLIR A65sc thermal camera is placed several centimeters in front of the \textit{M. asper funereus} elytra. Behind the elytra, a spherical mirror is placed. It enabled observation and measurement of both sides of one elytra of \textit{M. asper funereus}, the front side in direct light and the back side in reflected light (Fig.~\ref{fig:setup}). An elytron of \textit{M. asper funereus} is illuminated with several commercial laser beams with wavelengths at 405~nm, 450~nm, 532~nm, and 650~nm and with a declared output power of 5~mW. The laser was powered by standard AA batteries, so the real emitted power was usually less than the declared 5~mW. The FLIR A65sc thermal camera had an IR resolution of $640\times512$ pixels, an image frequency of 30~Hz, a detector type focal plane array (FPA), an uncooled VOX microbolometer, and a spectral range of 7.5–13~$\mathrm{\mu}$m. 

The recorded data were processed in the proprietary program "FLIR Research IR Max v4.40". For both the front and back sides, the average and maximum temperatures were calculated over non-overlapping $ 45\times45$-pixel areas at 18 frames per second for 30 seconds.

\subsection{Image Analysis}
Two-dimensional fast Fourier transforms (2D-FFTs), implemented in the Gwyddion software package, are computed from scanning probe images. In all related figures, the absolute value of the complex Fourier coefficient is shown, proportional to the square root of the power spectrum density. 

\section{Results and Discussion}

Several of the natural photonic structures studied here have previously been identified as promising platforms for radiative light control and materials engineering applications~\cite{Vasiljevic2021,Pavlovic2023} \textit{M. asper funereus} has served as a model system for investigating the influence of natural photonic architectures on thermal regulation. Its elytra consists of black and gray areas, the emissivity of which was the same, as measured by optical methods~\cite{Vasiljevic2021,Pavlovic2023}. This is because the wavelength of thermal radiation is close to the characteristic dimensions of the structure (hair-shaped structure) that covers the black and gray zones of the elytra. Both types of hair efficiently scatter radiation, as observed in other insects, including ants~\cite{Schwind2024}, thereby increasing the probability of absorbing radiation. Beyond optical properties, the elytra also show surprising thermodynamic behavior. Here, we present the first measurements of the thermodynamic response of these integuments, obtained from heat-capacity measurements over a temperature range. In addition, we used a thermal camera to monitor the temperature gradient distribution induced by irradiating beetle elytra with laser light at different wavelengths. The insect cuticle is primarily composed of chitin-protein complexes, and chitin is the second most abundant polysaccharide in nature after cellulose~\cite{Gilbert,Verstraete2019}.
The DSC response of pure chitin films has already been measured, and the monitored heat capacity monotonically increases with increasing temperature~\cite{Wen2007HeatCapacities, Guinesi2006DSC, Toffey1996ChitinKinetics, Kim1994ThermalChitin}. This monotonic increase is similar for almost all reported films and bulk materials. 

However, non-monotonic behavior has been reported in the literature for steels, some fluids, and polymers, particularly in association with glass and phase transitions~\cite{righetti2017crystallization, wurm2012crystallization, jariyavidyanont2021kinetics, he2018comparing}. In such cases, the heat capacity may exhibit transient negative dips during the transformation process. We emphasize that the behavior observed here is fundamentally different from these previously reported phenomena. Our measurements were performed over a relatively narrow temperature range (20 to 90°C), within which chitin does not undergo any known phase transition. Therefore, the decrease in the apparent heat capacity cannot be attributed to phase-transition effects or to changes in the material's intrinsic chemical properties. Instead, it arises from the development of surface corrugation. Corrugation modifies the system geometry and can influence the effective refractive index, curvature, wetting properties, and chemical potential~\cite{kammer2025theoretical}, thereby altering the system's thermodynamic response.

\begin{figure}
\centering
\includegraphics[width=1\textwidth]{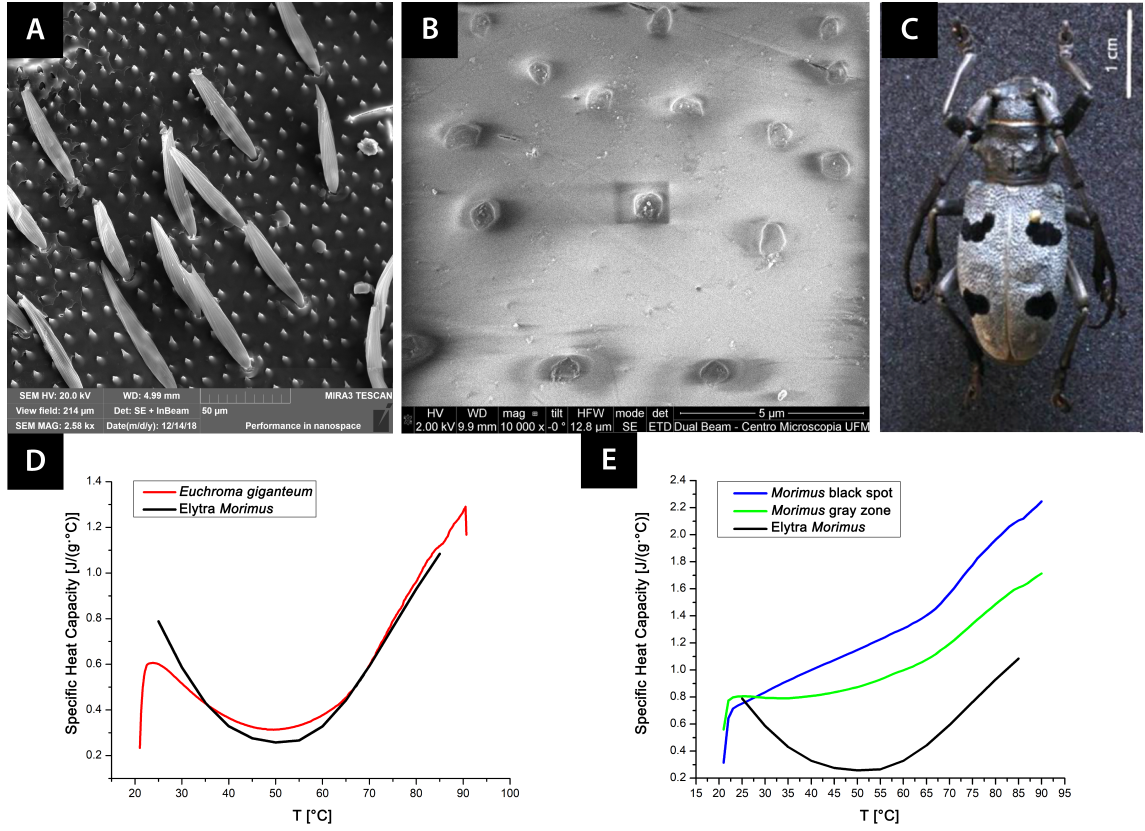}
\caption{\label{fig:Cp_pos} {Species‑specific surface patterns on the elytra of \textit{Morimus asper funereus} (A) and \textit{Euchroma giganteum} (B) reveal how geometry observed here by SEM may affect the thermal behavior. The heat capacity of the entire elytra of \textit{M. asper funereus} (C) and \textit{E. giganteum} exhibits minima as a function of the temperature (D). This is a behavior distinct from reported films and bulk materials~\cite{Wen2007HeatCapacities,Guinesi2006DSC,Toffey1996ChitinKinetics,Kim1994ThermalChitin}. The heat capacity of the black spots and the gray zone of \textit{M. asper funereus} also differs from the curve corresponding to an entire elytron. Figures (B) and (C) were reproduced from ref.~\cite{RINCONCELIS2015} and ref.~\cite{Vasiljevic2021}, respectively. }\label{fig:microstructure}}
\end{figure} 
It is interesting to see that both measured species exhibit a decrease in heat capacity over a narrow temperature range (Fig.~\ref{fig:microstructure}). The observed dependence indicates very efficient cooling in this temperature range, as was already proposed for the \textit{M. asper funereus}\cite{Vasiljevic2021}. Compared to previously published measurements on bulk or film‑like chitin~\cite{Wen2007HeatCapacities,Guinesi2006DSC,Toffey1996ChitinKinetics, Kim1994ThermalChitin}, which consistently show a monotonic increase of heat capacity with temperature, the behavior observed here differs significantly. We attribute this behavior to geometric corrugation. Notably, the geometric corrugation can be related to the thermal diffusivity $\alpha$~\cite{Lide2009CRC90} through the equation:

\begin{equation}
    \alpha = \frac{k}{\rho c_{\mathrm{p}}}, \label{eq:thermal_diffusivity}
\end{equation} 
where $k$, $\rho$, and $c_{\mathrm{p}}$ denote the thermal conductivity, the mass density, and the specific heat capacity at constant pressure, respectively. Because thermal diffusivity is inversely proportional to $C_{\mathrm{p}}$, any geometry‑induced decrease in heat capacity directly enhances thermal diffusion. Despite the relatively low thermal conductivity of chitin (0.73-0.82~W/m·K), a reduction in heat capacity increases the diffusivity and thus the rate at which temperature perturbations relax.  Thermal diffusivity quantifies a material's ability to dissipate temperature gradients toward thermal equilibrium. 

Therefore, corrugated architectures with rich interfaces affect thermal transport by scattering thermal radiation and altering heat-flow pathways, ultimately increasing thermal energy dissipation. As the elytra of \textit{M. asper funereus} comprises a gray area and four black patches with different microstructures, we separately examined the heat capacity of these regions.
From Figure~\ref{fig:microstructure}E, it is clear that the gray zone structure affects heat capacity more than the black but less than the whole elytra. The results show that synergy between neighboring areas is responsible for controlling the thermal response. The differences in the thermodynamic response of black and gray areas cannot be revealed by using only optical methods~\cite{Vasiljevic2021}.
To extend our analysis, we examined how temperature gradients develop in \textit{M. asper funereus} elytra following laser illumination.

\begin{figure}
\centering
\includegraphics[width=1\textwidth]{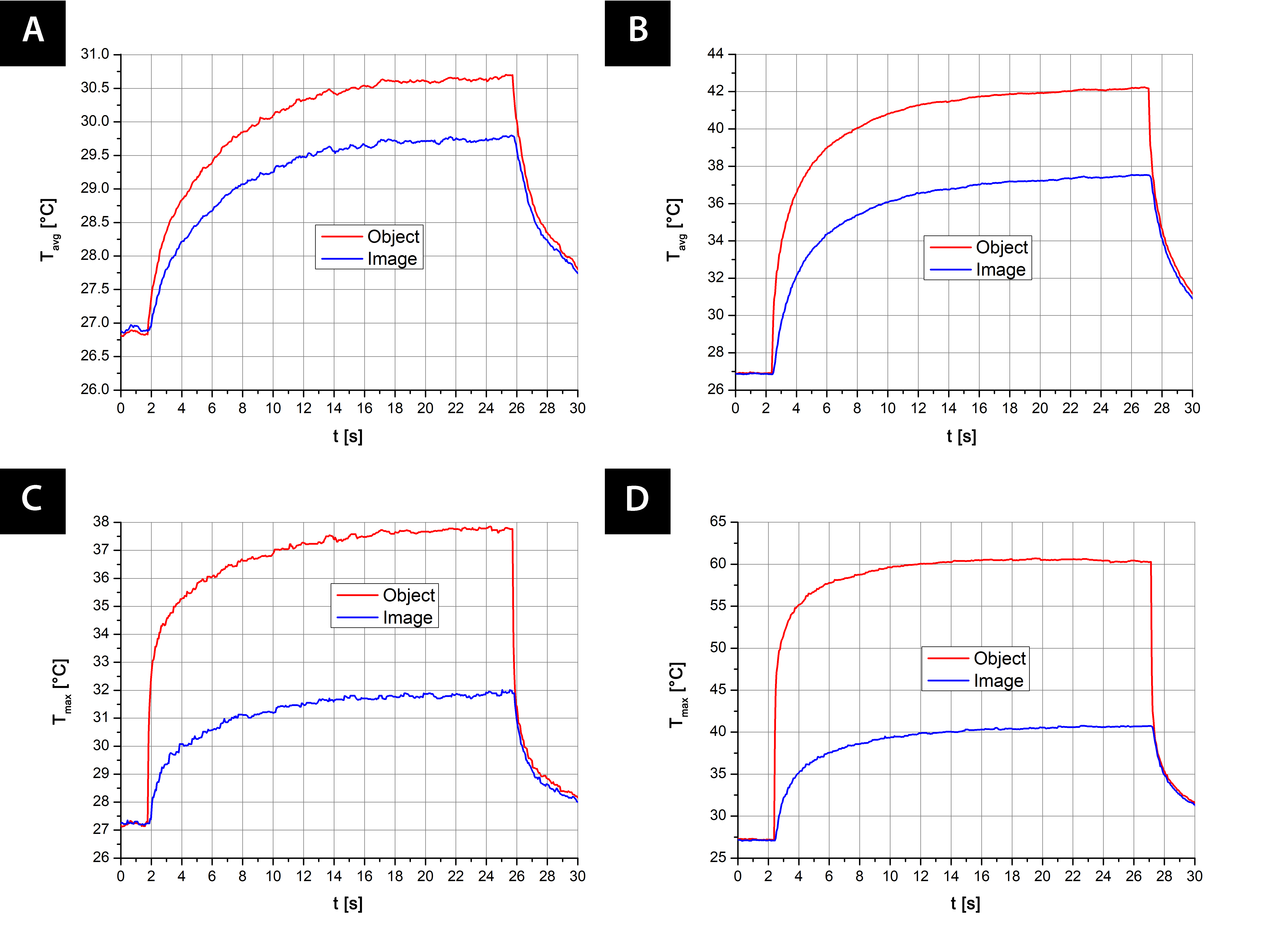}
\caption{Time‑resolved thermal imaging reveals how laser irradiation generates distinct front–back temperature gradients across the elytron in four measurements on \textit{Morimus asper funereus}. Average temperature of the outer (object) and inner (image) side of the elytra illuminated by a laser pointer at A) 405~nm, B) 650~nm. Maximum temperature of the outer (object) and inner (image) side of the elytra illuminated by a laser pointer at C) 405~nm, D) 650~nm. Here, "object" refers to the directly illuminated front side of the elytron, and "image" refers to the temperature of the back side as seen via reflection in the spherical mirror (Fig.~\ref{fig:setup}).
\label{fig:temperature_vs_time}}
\end{figure}

Considering the thickness of the elytra (a few microns) and the thermal coefficient of organic molecules, we expected a much smaller gradient and a rapid relaxation of temperature differences.
However, elytra are not planar, and their micro‑corrugation strongly influences radiative and conductive heat transport. The results presented in Figure~\ref{fig:temperature_vs_time} strongly confirm previous published results on \textit{M. asper funereus's} unusual thermal management~\cite{Vasiljevic2021} and are consistent with the measurements of heat capacity. The results presented in Figures~\ref{fig:Cp_pos} and \ref{fig:temperature_vs_time} show the link between corrugations and the unusual thermal management of \textit{M. asper funereus} and \textit{E. giganteum}. Furthermore, the observed heat capacity shows a similar pattern across different species, providing additional proof that chemical composition alone cannot account for the observed thermodynamic response.
\subsection{Theoretical study of apparent heat capacity from internal relaxation}
In a DSC experiment, the sample is driven through a prescribed temperature ramp, and the instrument records the external heat flow required to maintain the sample's temperature~\cite{hohne2003differential}. If, during the same scan, the material releases energy stored in metastable structural states, the measured heat flow is no longer solely due to ordinary warming~\cite{tool1946relation, narayanaswamy1971model, moynihan1976structural}. Part of the energy needed to raise the temperature is then supplied internally by the sample itself. We show below that sufficiently strong internal release can therefore produce an apparent negative heat-capacity signal without requiring the material's equilibrium heat capacity to be negative~\cite{callen1985thermodynamics, LyndenBell1977, Schmidt2001}. We describe this mechanism with a minimal energy balance. The imposed temperature ramp is
\begin{equation}
T(t)=T_0+rt,
\end{equation}
where \(r=dT/dt\) is the heating rate. In the absence of internal relaxation, increasing the sample temperature by \(dT\) requires an amount of heat per unit mass \(c_p^{(0)}(T)dT\). Dividing by time gives the external power per unit mass required by the calorimeter,
\begin{equation}
p_0(T)=c_p^{(0)}(T)\,r,
\end{equation}
where \(c_p^{(0)}(T)>0\) is the ordinary background heat capacity. This term represents the sensible heat required to warm the material at the imposed rate. If internal degrees of freedom release heat at a rate \(p_{\rm rel}(T)\) per unit mass, this heat contributes to the imposed warming and reduces the power that must be supplied externally. The calorimeter, therefore, measures the net external power
\begin{equation}
p_{\rm DSC}(T)
=
c_p^{(0)}(T)\,r
-
p_{\rm rel}(T).
\end{equation}
The minus sign follows directly from the energy balance. Heat released inside the sample partially replaces heat that would otherwise have to be delivered by the DSC. When \(p_{\rm rel}=0\), the usual positive response is recovered. When \(p_{\rm rel}\) becomes comparable to or larger than \(c_p^{(0)}r\), the measured heat-flow signal is strongly suppressed and may change sign. The apparent heat capacity inferred from the DSC signal is then
\begin{equation}
c_p^{\rm app}(T)
=
\frac{p_{\rm DSC}(T)}{r}
=
c_p^{(0)}(T)
-
\frac{p_{\rm rel}(T)}{r}.
\label{eq:cp_app_balance}
\end{equation}
Thus, the sign of \(c_p^{\rm app}\) is controlled by a competition between ordinary heat storage and internal heat release. The apparent heat capacity becomes negative over a finite interval whenever
\begin{equation}
p_{\rm rel}(T)>r\,c_p^{(0)}(T).
\label{eq:negative_cp_condition}
\end{equation}
\subsection{Activated relaxation of stored enthalpy}

\begin{figure}[t!]
    \centering
    \includegraphics[width=0.72\linewidth]{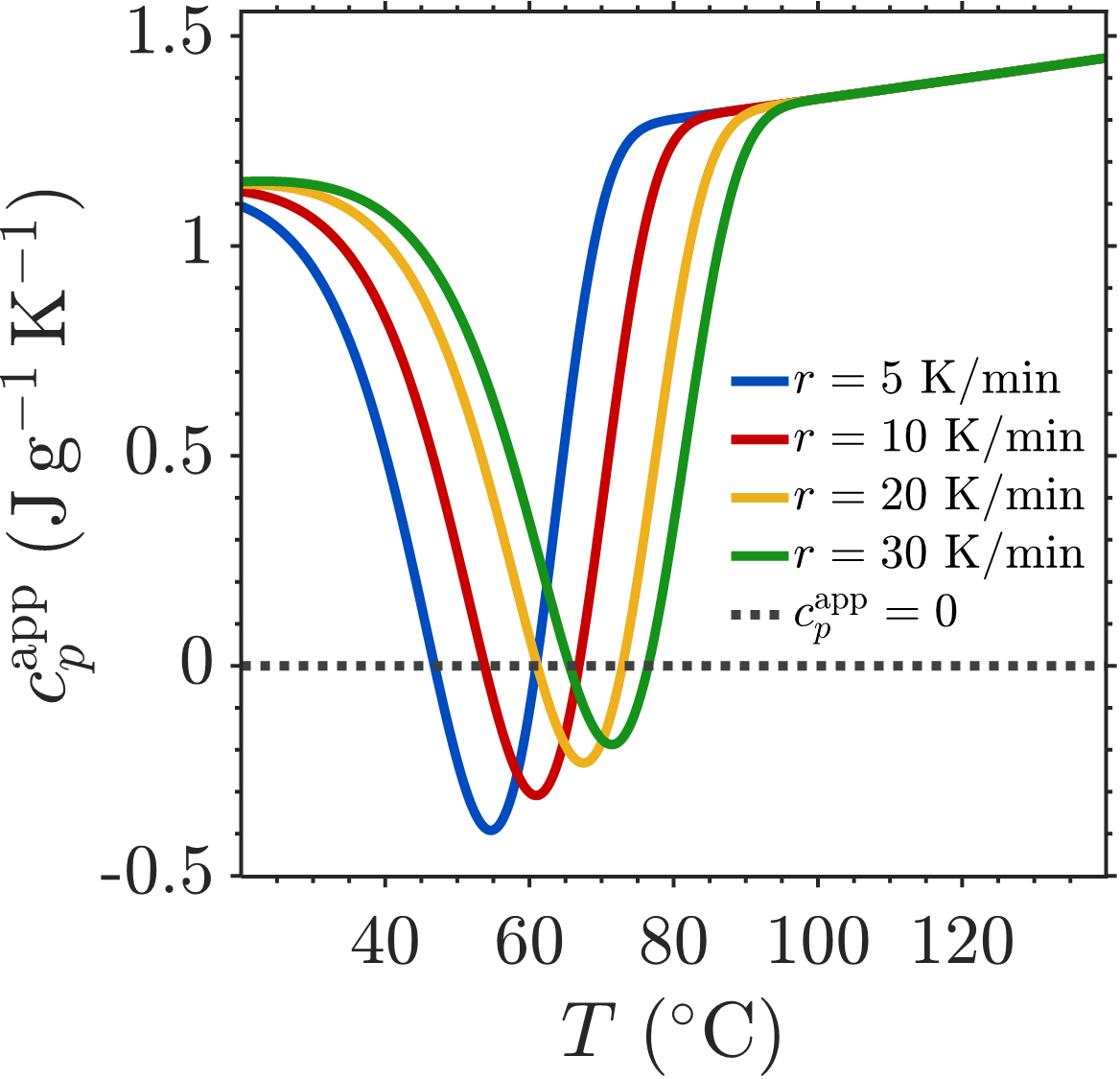}
    \caption{
    The model predicts that sufficiently strong internal relaxation may drive the apparent heat-capacity signal below \(0\rm~J/g/K\) under certain conditions.
    Heating-rate dependence of the apparent heat-capacity anomaly predicted by the internal-relaxation model.
    The sample is driven through a temperature ramp \(T(t)=T_0+rt\), while metastable internal states release stored enthalpy during the scan. Increasing the heating rate \(r\) shifts the \(c_p^{\rm app}\) dip to higher temperature because the internal states have less time to relax at a given temperature. The dotted horizontal line marks \(c_p^{\rm app}=0\rm~J/g/K\). This rate-dependent shift is a characteristic signature of thermally activated relaxation and provides an experimental test distinguishing an apparent DSC anomaly from a reversible equilibrium heat capacity.
    }
    \label{fig:Capp}
\end{figure}

To parameterize the heat-release term \(p_{\rm rel}\), we use a coarse-grained description in which the elytron contains local configurations that can relax during heating. These configurations are not meant to identify a unique microscopic process at this stage. They provide an effective representation of possible interfacial, mechanical, hydration-related, or structural rearrangements that store excess energy and release part of it once thermal activation becomes appreciable~\cite{moynihan1976structural}. Each family of relaxing configurations is assigned an activation barrier \(E_i\). This barrier is the energy scale that delays the rearrangement. At low temperatures, thermal fluctuations are usually insufficient to overcome it, whereas at higher temperatures, relaxation becomes more probable. We denote by \(n_i(T)\) the fraction of configurations of type \(i\) that remain unrelaxed at temperature \(T\), and by \(w_i\) their statistical weight, normalized as
\begin{equation}
\sum_i w_i=1 .
\end{equation}
The weights specify the relative contribution of each family to the ensemble of metastable states. The relaxation rate is taken in Arrhenius form,
\begin{equation}
\Gamma_i(T)
=
\Gamma_0
\exp\!\left[-\frac{E_i}{k_{\rm B}T}\right],
\end{equation}
where \(\Gamma_0\) is an attempt rate~\cite{laidler1984development}. A larger barrier requires a higher temperature, or a longer time, to relax appreciably. At fixed temperature, the unrelaxed fraction obeys
\begin{equation}
\frac{dn_i}{dt}
=
-\Gamma_i(T)n_i .
\end{equation}
During a DSC scan, temperature is the natural control variable. Using \(dT/dt=r\), the same relaxation law becomes
\begin{equation}
\frac{dn_i}{dT}
=
-\frac{\Gamma_i(T)}{r}n_i .
\label{eq:population_relaxation}
\end{equation}
The heating rate, therefore, controls the time available for relaxation at each temperature. A slower ramp allows the internal states to relax earlier, whereas a faster ramp shifts the response to higher temperature~\cite{kissinger1957reaction}. The heat released by these processes is determined by the decrease of stored enthalpy $h_{\rm st}$. Enthalpy is the relevant energy-like quantity because DSC measures heat flow during a temperature ramp at approximately fixed pressure~\cite{hohne2003differential}. If \(\Delta h_{\rm tot}\) is the total releasable enthalpy per unit mass when all metastable configurations relax, the remaining stored enthalpy is modeled as
\begin{equation}
h_{\rm st}(T)
=
\Delta h_{\rm tot}\sum_i w_i n_i(T).
\end{equation}
As the populations \(n_i\) decay, \(h_{\rm st}\) decreases. The released power per unit mass is therefore
\begin{equation}
p_{\rm rel}(T)
=
-\frac{dh_{\rm st}}{dt}.
\end{equation}
Combining this relation with Eq.~\eqref{eq:population_relaxation} gives
\begin{equation}
p_{\rm rel}(T)
=
\Delta h_{\rm tot}
\sum_i w_i \Gamma_i(T)n_i(T),
\end{equation}
which is positive because relaxation reduces the stored enthalpy and releases heat into the sample. Substitution into the DSC balance (Eq.~\ref{eq:cp_app_balance}) yields
\begin{equation}
c_p^{\rm app}(T)
=
c_p^{(0)}(T)
-
\frac{\Delta h_{\rm tot}}{r}
\sum_i w_i \Gamma_i(T)n_i(T).
\label{eq:cp_app_model}
\end{equation}
This expression is the working model for the apparent heat-capacity anomaly. The first term is the ordinary positive heat capacity. The second term is the reduction in externally supplied heat caused by internal heat release. Its magnitude is controlled by the stored enthalpy, the distribution of activation barriers, and the heating rate. Equation~\eqref{eq:cp_app_model} gives the minimal mechanism underlying the numerical curves plotted in Fig.~\ref{fig:Capp}.

\subsection{Geometry-dependent barrier spectrum}

The final step is to connect the effective relaxation spectrum to the elytron's morphology. In the present model, geometry does not directly cause the background heat capacity to be negative. Instead, it modifies the local environments in which metastable states form, thereby altering the activation barriers that govern their relaxation upon heating. A corrugated interface contains regions with different curvatures, strains, thicknesses, and local molecular environments. The relevance of curvature is already evident in the local chemical potential. For a surface with principal radii of curvature \(R_1\) and \(R_2\), one may write, in the spirit of Gibbs--Thomson-type interfacial thermodynamics (Eq.~\ref{eq:ChemicalPotential})~\cite{johnson1965generalization, perez2005gibbs},
where the curvature variable entering this thermodynamic shift is
\begin{equation}
\kappa
=
\frac{1}{R_1}
+
\frac{1}{R_2}.
\end{equation}
It is positive for convex regions, negative for concave regions, and vanishes for a locally flat surface. A strongly corrugated material, therefore, contains a distribution of local chemical potentials rather than a single uniform environment. The same geometrical heterogeneity can be incorporated phenomenologically into the activation barriers. For a local region \(i\), we write
\begin{equation}
E_i
=
E_0
+
\lambda_1\kappa_i
+
\lambda_2\kappa_i^2
+
\lambda_s s_i
+\cdots,
\qquad
\kappa_i=
\frac{1}{R_{1,i}}+\frac{1}{R_{2,i}}.
\label{eq:geometry_barrier}
\end{equation}
Here \(E_0\) is a reference barrier, while \(s_i\) denotes additional local structural variables such as strain, thickness, hydration, or interfacial state. The coefficients \(\lambda_1\), \(\lambda_2\), and \(\lambda_s\) are phenomenological parameters. They do not specify a microscopic chemistry; they only encode the possibility that relaxation barriers differ between flat, curved, strained, or locally hydrated regions. The relevant object is therefore not a single activation barrier, but a geometry-dependent distribution \(P_{\rm geom}(E)\). In continuum form, the apparent heat capacity becomes
\begin{equation}
c_p^{\rm app}(T)
=
c_p^{(0)}(T)
-
\frac{1}{r}
\int dE\,
P_{\rm geom}(E)\,
\epsilon(E)\,
\Gamma(E,T)\,
n(E,T),
\label{eq:continuous_cp_model}
\end{equation}
where \(\epsilon(E)\) is the enthalpy released by states with barrier \(E\), \(\Gamma(E,T)\) is their thermally activated relaxation rate, and \(n(E,T)\) is the remaining unrelaxed fraction. This expression summarizes the model's physical content. The surface architecture can shape the distribution of metastable environments; that distribution controls the temperature range, width, and amplitude of the heat-release anomaly; and the resulting heat release modifies the heat flow measured by DSC. 
The simple model used in our study qualitatively reproduces the shape of the recorded thermodynamic response with some $c_p^{\rm app}$ values even below $0\rm~J/g/K$. 
 The results presented suggest that more complex natural structures (with a more pronounced interfacial effect) could even exhibit an apparent negative heat capacity (Fig.~\ref{fig:Capp}). So far, negative heat capacity has been observed only in systems such as stellar gas, self-gravitating nonequilibrium systems, and nanoclusters~\cite{LyndenBell1977,LYNDENBELL1999,Katz2000,Schmidt2001,Posch2006,LyndenBell2008,Boksenbojm2011}. It is often explained by the local violation of the second law of thermodynamics~\cite{LyndenBell1977,LyndenBell2008}.
Recently, a theoretical study showed~\cite{OConnorRamgoolam2024Permutation} that a negative specific heat capacity can be linked to superexponential growth of degeneracies at low energies, $k\lesssim N\log{N}$. This is directly related to the presence of an interface, leading to the degeneracy by breaking symmetry, generating strain or potential, and splitting previously identical energy states. 
To further strengthen the link between the measured thermodynamic properties and geometry, we performed a detailed 2D FFT analysis of the corresponding SEM images.
\begin{figure}
\centering
\includegraphics[width=1\textwidth]{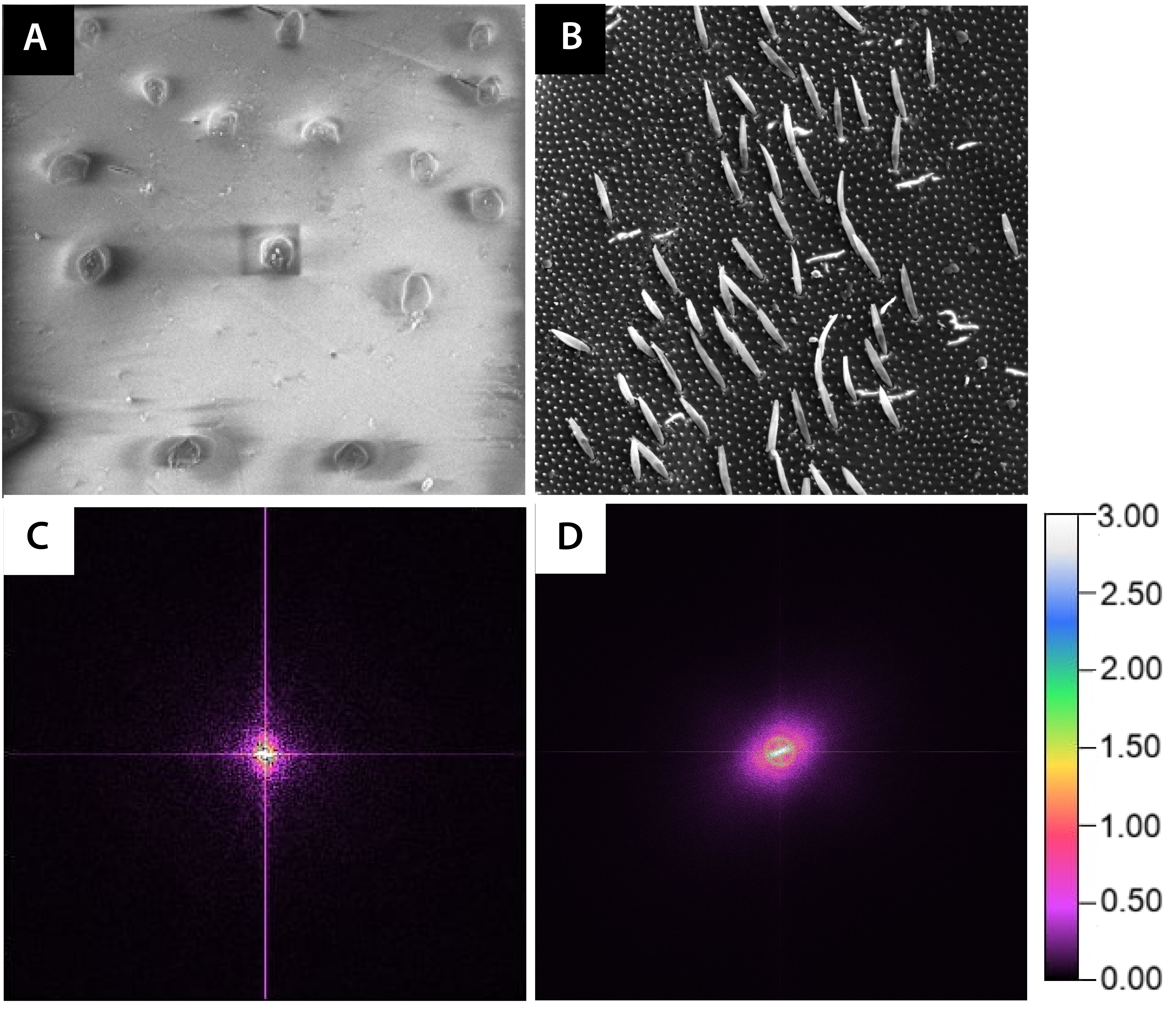}
\caption{\label{fig:Euchroma} SEM images of \textit{Euchroma giganteum} (a) and \textit{Morimus asper funereus} (b).  2D-FFT (c,d) obtained from SEM images Figure (a) was reproduced from ref.~\cite{RINCONCELIS2015}.
}
\end{figure}

The 2D-FFT analysis provides additional insight into the effects of ordering at the nano- and mesoscales. When the effect of geometry on  $c_{\mathrm{p}}$ is less profound, the 2D-FFT maps exhibit more isotropic frequency distributions, while stronger geometry effects will lead to the appearance of a more disk-shaped frequency profile.
The experimental studies described in this article, along with the theoretical model, clearly confirm an unusual yet robust effect of geometry on the thermodynamic response. Over the past two decades, the influence of geometry on optical response has been extensively investigated, both experimentally and numerically~\cite{Kinoshita2008,mouchet2018structural,Verstraete2019,Mouchet2021,Mara2022, hatifi2022b, hatifi2026}. These studies have demonstrated that structural organization can profoundly modify the interaction between matter and electromagnetic radiation. A related phenomenon is thermal emission (TE), a universal property of matter above absolute zero Kelvin that originates from the thermal motion of atoms and molecules. 

Vibrational excitations of chemical bonds generate fluctuating electromagnetic currents, which, in turn, produce thermal radiation in accordance with Planck's spectral distribution. Despite significant advances in energy-efficient materials, TE remains a major source of heat waste in many technological applications~\cite{Mouchet2025,Delmote2026}. Beyond their optical behavior, complex surfaces and interfaces can introduce additional degrees of freedom that modify energy-level degeneracy, local electromagnetic fields, and thermodynamic stability. Such observations suggest that structural design may play a broader role in governing thermal behavior than previously recognized. Understanding these interface-driven effects, therefore, opens new opportunities for tailoring thermal properties. Additionally, researchers have recently studied and published on how wrinkles (corrugation/curvature) in 2D materials (where the surface dominates over the bulk) affect electrical properties \cite{Iyengar}.

In particular, complex geometries capable of inducing negative apparent $c^{\rm app}_{\mathrm{p}}$ could provide a route to reduce TE and mitigate unwanted heat losses. Such concepts may prove especially valuable in applications requiring efficient thermal management, such as buildings or other large structures. More generally, the sophisticated corrugated structures found in natural photonic systems offer inspiration for developing next-generation energy-efficient materials and thermal management technologies.

\section*{Acknowledgments}
B. K., D. V., B. B., B. S., and M. R. R. acknowledge funding provided by the Institute of Physics Belgrade, through institutional funding by the Ministry of Education, Science, and Technological Development of the Republic of Serbia. B. K. also acknowledges financial support from FRS-FNRS. S. R. M. is a Research Associate of the Belgian Fund for Scientific Research (FRS-FNRS) and was supported by a Marie Skłodowska-Curie Actions Fellowship in the framework of BEWARE programme (Convention n°2110034) of the Walloon Region (COFUND Marie Skłodowska-Curie Actions of the European Union \#847587). B. B., M. H., D. V., B. K., T. V., M. R. R., and B. R. acknowledge the support of the EU: the EIC Pathfinder Challenges 2022 call through the Research Grant 101115149 (project ARTEMIS). The content
reflects only the authors’ views, and the European Commission is not responsible for any use that may be made
of the information it contains.

The supporting bodies did not play a role in the design of the study, the collection of samples, the analysis \& interpretation of the data, and the writing of the manuscript.

\section*{Author Contributions}

B. K., S. R. M., M. H., and T. V.: Conception, Supervision, Data Analysis, Interpretation, Discussion, Written manuscript, and Revision. M. H.: Theoretical Modeling, Analytical Calculations, and Numerical Simulations. B. B.: Data analysis, Discussion, Written manuscript, Visualization, and Revision. B. S., D. V., and S. O.: Experimental measurements and Discussion. N. R.: Data analysis, Discussion, and Revision. B. R., M. R. R.: Fourier Image Analysis, Discussion, and Revision.

\section*{Conflicts of interest}

The authors declare no conflict of interest.

\clearpage
\bibliographystyle{unsrt} 

\begin{thebibliography}{10}

\bibitem{krumm2017thermodynamics}
Marius Krumm, Howard Barnum, Jonathan Barrett, and Markus~P. M{\"u}ller.
\newblock Thermodynamics and the structure of quantum theory.
\newblock {\em New Journal of Physics}, 19(043025):043025, 2017.

\bibitem{gonzalez2012linear}
Luis Gonz{\'a}lez-Urbina, Kasper Baert, Branko Kolaric, Javier
  P{\'e}rez-Moreno, and Koen Clays.
\newblock Linear and nonlinear optical properties of colloidal photonic
  crystals.
\newblock {\em Chemical Reviews}, 112(4):2268--2285, 2012.

\bibitem{MADANU2023290}
Thomas~L. Madanu, S{\'e}bastien~R. Mouchet, Olivier Deparis, Jing Liu, Yu~Li,
  and Bao-Lian Su.
\newblock Tuning and transferring slow photons from {{TiO}}{$_2$} photonic
  crystals to {{BiVO}}{$_4$} nanoparticles for unprecedented visible light
  photocatalysis.
\newblock {\em Journal of Colloid and Interface Science}, 634:290--299, 2023.

\bibitem{LOURDUMADANU2023233}
Thomas~Lourdu Madanu, Laroussi Chaabane, S{\'e}bastien~R. Mouchet, Olivier
  Deparis, and Bao-Lian Su.
\newblock Manipulating multi-spectral slow photons in bilayer inverse opal
  {{TiO}}{$_2$}@{{BiVO}}{$_4$} composites for highly enhanced visible light
  photocatalysis.
\newblock {\em Journal of Colloid and Interface Science}, 647:233--245, 2023.

\bibitem{huang2024colloidal}
Yaxin Huang, Changjin Wu, Jingyuan Chen, and Jinyao Tang.
\newblock Colloidal self-assembly: {{From}} passive to active systems.
\newblock {\em Angewandte Chemie International Edition}, 63(9):e202313885,
  2024.

\bibitem{dellieu2015quantum}
Louis Dellieu, Olivier Deparis, J{\'e}r{\^o}me Muller, Branko Kolaric, and
  Micha{\"e}l Sarrazin.
\newblock Quantum vacuum photon modes and repulsive {{Lifshitz}}--van der
  {{Waals}} interactions.
\newblock {\em Physical Review B}, 92:235418, 2015.

\bibitem{israelachvili2011intermolecular}
Jacob~N. Israelachvili.
\newblock {\em Intermolecular and Surface Forces}.
\newblock Academic Press, London, UK, 3rd edition, 2011.

\bibitem{butt2003physics}
Hans-J{\"u}rgen Butt, Karlheinz Graf, and Michael Kappl.
\newblock {\em Physics and Chemistry of Interfaces}.
\newblock John Wiley \& Sons, Weinheim, Germany, 1 edition, 2003.

\bibitem{deGennes1985wetting}
Pierre-Gilles {de Gennes}.
\newblock Wetting: Statics and dynamics.
\newblock {\em Reviews of Modern Physics}, 57(3):827--863, 1985.

\bibitem{alhambra2014casimir}
{\'A}lvaro~M. Alhambra, Achim Kempf, and Eduardo {Mart{\'i}n-Mart{\'i}nez}.
\newblock Casimir forces on atoms in optical cavities.
\newblock {\em Physical Review A}, 89:033835, 2014.

\bibitem{biro2003}
L{\'a}szl{\'o}~P{\'e}ter Bir{\'o}, Zs~B{\'a}lint, K~Kert{\'e}sz, Z~V{\'e}rtesy,
  {\relax GI}~M{\'a}rk, {\relax ZE}~Horv{\'a}th, J~Bal{\'a}zs, D~M{\'e}hn,
  I~Kiricsi, V~Lousse, et~al.
\newblock Role of photonic-crystal-type structures in the thermal regulation of
  a {{Lycaenid}} butterfly sister species pair.
\newblock {\em Physical Review E}, 67(2):021907, 2003.

\bibitem{Carrascal2017}
Luis~M Carrascal, Yolanda~Jim{\'e}nez Ruiz, and Jorge~M Lobo.
\newblock Beetle exoskeleton may facilitate body heat acting differentially
  across the electromagnetic spectrum.
\newblock {\em Physiological and Biochemical Zoology}, 90(3):338--347, 2017.

\bibitem{Mouchet2025}
S{\'e}bastien~R. Mouchet.
\newblock Infrared absorbers inspired by nature.
\newblock {\em Journal of The Royal Society Interface}, 22(223):20240284, 2025.

\bibitem{fultz2016phase}
Brent Fultz.
\newblock Phase transitions in materials.
\newblock {\em American Journal of Physics}, 84(4):317--318, 2016.

\bibitem{Davies2020}
Paul C.~W. Davies.
\newblock Does new physics lurk inside living matter?
\newblock {\em Physics Today}, 73(8):34--40, 2020.

\bibitem{Bravetti2019Contact}
Alessandro Bravetti.
\newblock Contact geometry and thermodynamics.
\newblock {\em International Journal of Geometric Methods in Modern Physics},
  16(supp01):1940003, 2019.

\bibitem{mouchet2018structural}
S{\'e}bastien~R. Mouchet and Pete Vukusic.
\newblock Structural colours in lepidopteran scales.
\newblock In {\em Advances in Insect Physiology}, volume~54, pages 1--53.
  Elsevier, 2018.

\bibitem{Mouchet2021}
S{\'e}bastien~R. Mouchet and Olivier Deparis.
\newblock {\em Natural Photonics and Bioinspiration}.
\newblock Artech House, 2021.

\bibitem{Verstraete2019}
Charlotte Verstraete, S{\'e}bastien~R. Mouchet, Thierry Verbiest, and Branko
  Kolaric.
\newblock Linear and nonlinear optical effects in biophotonic structures using
  classical and nonclassical light.
\newblock {\em Journal of Biophotonics}, 12(1):e201800262, 2019.

\bibitem{Mara2022}
Dimitrije Mara, Bojana Bokic, Thierry Verbiest, S{\'e}bastien~R. Mouchet, and
  Branko Kolaric.
\newblock Revealing the wonder of natural photonics by nonlinear optics.
\newblock {\em Biomimetics}, 7(153), 2022.

\bibitem{Delmote2026}
Kevin Delmote, Amandine Marchand, Olivier Deparis, and S{\'e}bastien~R.
  Mouchet.
\newblock Bioinspired approach to infrared harvesting.
\newblock {\em Nanophotonics XI}, 14076:1407616, 2026.

\bibitem{Barron1999HeatCapacity}
T.~H.~K. Barron and G.~K. White.
\newblock {\em Heat Capacity and Thermal Expansion at Low Temperatures}.
\newblock International Cryogenics Monograph Series. Springer, New York, 1999.

\bibitem{Vasiljevic2021}
Darko Vasiljevi{\'c}, Danica Pavlovi{\'c}, Vladimir Lazovi{\'c}, Branko
  Kolari{\'c}, Branislav Salati{\'c}, Wang Zhang, Di~Zhang, and Dejan
  Panteli{\'c}.
\newblock Thermal radiation management by natural photonic structures:
  {{Morimus}} asper funereus case.
\newblock {\em Journal of Thermal Biology}, 98:102932, May 2021.

\bibitem{Pavlovic2023}
Marina~Simovi{\'c} Pavlovi{\'c}, Bojana Boki{\'c}, Charlotte Verstraete, Darko
  Vasiljevi{\'c}, S{\'e}bastien~R. Mouchet, Thierry Verbiest, and Branko
  Kolari{\'c}.
\newblock Holographic and nonlinear optical study of natural photonic
  structures: {{Where}} biology meets physics.
\newblock In {\em 2023 23rd International Conference on Transparent Optical
  Networks ({{ICTON}})}, pages 1--4, 2023.

\bibitem{Schwind2024}
Bertram Schwind, Xia Wu, Michael Tiemann, and Helge-Otto Fabritius.
\newblock Natural near field coupled leaky-mode resonant anti-reflection
  structures: The setae of {{Cataglyphis}} bombycina.
\newblock {\em Frontiers in Physics}, Volume 12 - 2024, 2024.

\bibitem{Gilbert}
Lawrence~I. Gilbert.
\newblock {\em Insect Development: Morphogenesis, Molting and Metamorphosis}.
\newblock Academic Press, Amsterdam, The Netherlands, 2009.

\bibitem{Wen2007HeatCapacities}
Jianye Wen.
\newblock Heat capacities of polymers.
\newblock In James~E. Mark, editor, {\em Physical Properties of Polymers
  Handbook}, pages 145--154. Springer, New York, NY, 2007.

\bibitem{Guinesi2006DSC}
Luciana~Simionatto Guinesi and {\'E}der T.~G. Cavalheiro.
\newblock The use of {{DSC}} curves to determine the acetylation degree of
  chitin/chitosan samples.
\newblock {\em Thermochimica Acta}, 444(2):128--133, 2006.

\bibitem{Toffey1996ChitinKinetics}
A.~Toffey, G.~Samaranayake, C.~E. Frazier, and W.~G. Glasser.
\newblock Chitin derivatives. {{I}}. {{Kinetics}} of the heat-induced
  conversion of chitosan to chitin.
\newblock {\em Journal of Applied Polymer Science}, 60(1):75--85, 1996.

\bibitem{Kim1994ThermalChitin}
Seong~Soo Kim, Seon~Jeong Kim, Yoon~Duk Moon, and Young~Moo Lee.
\newblock Thermal characteristics of chitin and hydroxypropyl chitin.
\newblock {\em Polymer}, 35(15):3212--3216, July 1994.

\bibitem{righetti2017crystallization}
Maria~Cristina Righetti.
\newblock Crystallization of polymers investigated by temperature-modulated
  {{DSC}}.
\newblock {\em Materials}, 10(4):442, 2017.

\bibitem{wurm2012crystallization}
Andreas Wurm, Evgeny Zhuravlev, Kathrin Eckstein, Dieter Jehnichen, Doris
  Pospiech, R.~Androsch, B.~Wunderlich, and Christoph Schick.
\newblock Crystallization and homogeneous nucleation kinetics of
  poly({{$\varepsilon$}}-caprolactone) ({{PCL}}) with different molar masses.
\newblock {\em Macromolecules}, 45(9):3816--3828, 2012.

\bibitem{jariyavidyanont2021kinetics}
Katalee Jariyavidyanont, Evgeny Zhuravlev, Christoph Schick, and Ren{\'e}
  Androsch.
\newblock Kinetics of homogeneous crystal nucleation of polyamide 11 near the
  glass transition temperature.
\newblock {\em Polymer Crystallization}, 4(1):e10149, 2021.

\bibitem{he2018comparing}
Yucheng He, Ruiqi Luo, Zhaolei Li, Ruihua Lv, Dongshan Zhou, Soonho Lim,
  Xiaoning Ren, Hongxu Gao, and Wenbing Hu.
\newblock Comparing crystallization kinetics between polyamide 6 and polyketone
  via chip-calorimeter measurement.
\newblock {\em Macromolecular Chemistry and Physics}, 219(3):1700385, 2018.

\bibitem{kammer2025theoretical}
Michael~N. Kammer, Amanda~K. Kussrow, and Darryl~J. Bornhop.
\newblock Theoretical basis for refractive index changes resulting from
  solution phase molecular interaction.
\newblock {\em The Journal of Physical Chemistry B}, 129(13):3297--3305, 2025.

\bibitem{RINCONCELIS2015}
Ra{\'u}l~L. {Rinc{\'o}n-Celis}, Diego {Bernal-Garc{\'i}}, Herbert
  {Vinck-Posada}, and Gabriel Colorado.
\newblock {Simulaci\'on en FDFD para describir el fen\'omeno de iridiscencia en
  los \'elitros del {{\emph{Euchroma gigantea}}}}.
\newblock {\em Momento}, pages 16--30, December 2015.

\bibitem{Lide2009CRC90}
David~R. Lide, editor.
\newblock {\em {{CRC}} Handbook of Chemistry and Physics: A Ready-Reference
  Book of Chemical and Physical Data}.
\newblock CRC Press, Boca Raton, FL, 90 edition, 2009.

\bibitem{hohne2003differential}
G.~W.~H. H{\"o}hne, W.~F. Hemminger, and H.-J. Flammersheim.
\newblock {\em Differential Scanning Calorimetry}.
\newblock Springer, Berlin, Heidelberg, 2 edition, 2003.

\bibitem{tool1946relation}
Arthur~Q. Tool.
\newblock Relation between inelastic deformability and thermal expansion of
  glass in its annealing range.
\newblock {\em Journal of the American Ceramic Society}, 29(9):240--253, 1946.

\bibitem{narayanaswamy1971model}
O.~S. Narayanaswamy.
\newblock A model of structural relaxation in glass.
\newblock {\em Journal of the American Ceramic Society}, 54(10):491--498, 1971.

\bibitem{moynihan1976structural}
C.~T. Moynihan, P.~B. Macedo, C.~J. Montrose, P.~K. Gupta, M.~A. DeBolt, J.~F.
  Dill, B.~E. Dom, P.~W. Drake, A.~J. Easteal, P.~B. Elterman, R.~P. Moeller,
  H.~Sasabe, and J.~A. Wilder.
\newblock Structural relaxation in vitreous materials.
\newblock {\em Annals of the New York Academy of Sciences}, 279(1):15--35,
  1976.

\bibitem{callen1985thermodynamics}
Herbert~B. Callen.
\newblock {\em Thermodynamics and an Introduction to Thermostatistics}.
\newblock John Wiley \& Sons, New York, 2 edition, 1985.

\bibitem{LyndenBell1977}
D.~{Lynden-Bell} and R.~M. {Lynden-Bell}.
\newblock On the negative specific heat paradox.
\newblock {\em Monthly Notices of the Royal Astronomical Society},
  181(3):405--419, December 1977.

\bibitem{Schmidt2001}
Martin Schmidt, Robert Kusche, Thomas Hippler, J{\"o}rn Donges, Werner
  Kronm{\"u}ller, Bernd {von Issendorff}, and Hellmut Haberland.
\newblock Negative heat capacity for a cluster of 147 sodium atoms.
\newblock {\em Physical Review Letters}, 86(7):1191--1194, 2001.

\bibitem{laidler1984development}
Keith~J. Laidler.
\newblock The development of the arrhenius equation.
\newblock {\em Journal of Chemical Education}, 61(6):494, 1984.

\bibitem{kissinger1957reaction}
H.~E. Kissinger.
\newblock Reaction kinetics in differential thermal analysis.
\newblock {\em Analytical Chemistry}, 29(11):1702--1706, 1957.

\bibitem{johnson1965generalization}
Charles~A. Johnson.
\newblock Generalization of the {{Gibbs-Thomson}} equation.
\newblock {\em Surface Science}, 3(5):429--444, 1965.

\bibitem{perez2005gibbs}
Michel Perez.
\newblock Gibbs-{{Thomson}} effects in phase transformations.
\newblock {\em Scripta Materialia}, 52(8):709--712, 2005.

\bibitem{LYNDENBELL1999}
D.~{Lynden-Bell}.
\newblock Negative specific heat in astronomy, physics and chemistry.
\newblock {\em Physica A: Statistical Mechanics and its Applications},
  263(1):293--304, 1999.

\bibitem{Katz2000}
Joseph Katz and Isao Okamoto.
\newblock Fluctuations in isothermal spheres.
\newblock {\em Monthly Notices of the Royal Astronomical Society},
  317(1):163--169, September 2000.

\bibitem{Posch2006}
Harald~A. Posch and Walter Thirring.
\newblock Thermodynamic instability of a confined gas.
\newblock {\em Physical Review E: Statistical Physics, Plasmas, Fluids, and
  Related Interdisciplinary Topics}, 74(5):051103, 2006.

\bibitem{LyndenBell2008}
D.~{Lynden-Bell} and R.~M. {Lynden-Bell}.
\newblock Negative heat capacities do occur. {{Comment}} on ``{{Critical}}
  analysis of negative heat capacities in nanoclusters'' by {{Michaelian K}}.
  and {{Santamar\'ia-Holek I}}.
\newblock {\em Europhysics Letters}, 82(4):43001, 2008.

\bibitem{Boksenbojm2011}
E.~Boksenbojm, C.~Maes, K.~Neto{\v c}n{\'y}, and J.~Pe{\v s}ek.
\newblock Heat capacity in nonequilibrium steady states.
\newblock {\em Europhysics Letters}, 96(4):40001, 2011.

\bibitem{OConnorRamgoolam2024Permutation}
Denjoe O'Connor and Sanjaye Ramgoolam.
\newblock Permutation invariant matrix quantum thermodynamics and negative
  specific heat capacities in large {{N}} systems.
\newblock {\em Journal of High Energy Physics}, 2024(12):161, December 2024.

\bibitem{Kinoshita2008}
Shuichi Kinoshita.
\newblock {\em Structural Colors in the Realm of Nature}.
\newblock World Scientific, Singapore, 2008.

\bibitem{hatifi2022b}
Mohamed Hatifi, Dimitrije Mara, Bojana Bokic, Rik Van~Deun, Brian Stout,
  Emmanuel Lassalle, Branko Kolaric, and Thomas Durt.
\newblock Fluorimetry in the {{Strong-Coupling Regime}}: {{From}} a
  {{Fundamental Perspective}} to {{Engineering New Tools}} for {{Tracing}} and
  {{Marking Materials}} and {{Objects}}.
\newblock {\em Applied Sciences}, 12(18):9238, September 2022.

\bibitem{hatifi2026}
Mohamed Hatifi.
\newblock Geometry-controlled freezing and revival of {{Bell}} nonlocality
  through environmental memory.
\newblock {\em Physical Review A}, 113(2):022204, February 2026.

\bibitem{Iyengar}
Sathvik~Ajay Iyengar, James~G. McHugh, Jonathan~P. Salvage, Robert Vajtai,
  Venkataramana Gadhamshetty, Alan~B. Dalton, Manoj Tripathi, Pulickel~M.
  Ajayan, and Vincent Meunier.
\newblock Sub-{{Nanometer Curvature Unlocks Quantum Orbital Flexoelectricity}}
  in {{Graphene}}.
\newblock {\em Advanced Materials}, page e18224, July 2026.

\end{thebibliography}

\end{document}